\documentclass[preprintnumbers,amsmath,amssymb]{revtex4}

\usepackage{graphicx}
\usepackage{dcolumn}
\usepackage{bm}

\usepackage{amssymb}

\newcommand{\be}{\begin{equation}}
\newcommand{\ee}{\end{equation}}
\newcommand{\ba}{\begin{eqnarray}}
\newcommand{\ea}{\end{eqnarray}}
\newcommand{\rf}[1]{(\ref{#1})}
\newcommand{\bi}{\bibitem}

\newcommand{\bel}[1]{\begin{equation}\label{#1}}
\newcommand{\qe}{\end{equation}}
\begin{document}
\title{The tragedy of the commons in a multi-population complementarity game}
\author{Wei Li}
\affiliation{Complexity Science Center, Hua-Zhong Normal
University, \\Wuhan 430079, P.R. China}
\author{J\"urgen Jost}
\affiliation{Max Planck Institute for Mathematics in the
Sciences,\\ Inselstr.22, 04103 Leipzig, Germany}

\begin{abstract}
  We study a complementarity game with multiple populations whose
  members' offered contributions are  put together towards some
  common aim. When the sum of the players' offers reaches or exceeds
  some threshold $K$, they each receive $K$ minus their own
  offers. Else, they all receive nothing. Each player tries to offer as
  little as possible, hoping that the sum of the contributions still
  reaches $K$, however. The game is symmetric at the individual level,
  but has many equilibria that are more or less favorable to the
  members of certain populations. In particular, it is possible that
  the members of one or several populations do not contribute
  anything, a behavior called defecting, while the others still
  contribute enough to reach the threshold. Which of these equilibria then is
  attained is decided by the dynamics at the population level that in
  turn depends on the strategic options the players possess. We find
  that defecting occurs when more than 3 populations participate in
  the game, even when the strategy scheme employed is very simple, if
  certain  conditions for the system parameters are satisfied. The
  results are obtained through systematic simulations.
\end{abstract}
\maketitle

\section{Introduction}

We consider a complementarity game with several populations that
need to cooperate to achieve some common goal. In each round,  by
random assignment, groups of $I$ individuals are formed, one from
each population. Here, $I$ of course is the number of
participating populations. Each agent $i$ offers to contribute
$k_i$ units.\footnote{All numbers will be nonnegative integers
$\le K$.} In order to reach the common goal, at least $K$  units
have to be provided. The question then is which of the  partners
should contribute how much to reach that level $K$. Thus, the
contributions should satisfy \bel{1} \sum_{i} k_i\ge K. \qe The
pay-off for agent $i$ then is \bel{2} K-k_i. \qe The pay-off is 0
when \rf{1} is not satisfied. Thus, in order to maximize her/his
pay-off, each agent wants to contribute as little as possible,
provided however, that the joint contribution of the two partners
satisfies \rf{1}. In particular, in this game, any collection of
values $k_i$ with \bel{3} \sum_i k_i= K \qe yield a Nash
equilibrium. This means that neither agent can gain by
unilaterally deviating from it when the other player does not
change her/his contribution. Increasing the own contribution would
simply be wasteful, as it lowers the pay-off \rf{2},
and decreasing it would make the entire endeavor fail, by violating \rf{1}, and the pay-off would shrink to 0.\\
Thus, there are multiple Nash equilibria. The question then is
which one of them is attained. The fairest equilibrium seems to be
\bel{4} k_i=\frac{K}{I}, \qe that is, where the partners
contribute equally.\footnote{Of course,
  $\frac{K}{I}$ need not be an integer, so that, more precisely, $k_i$
  should be smallest integer $\ge \frac{K}{I}$, but we shall choose
  $K$ and ignore this trivial technical point.} There is nothing that
guarantees, however, that such a fair situation will be achieved
in groups of selfish players. When the players can negotiate, the
outcome will depend on their respective negotiating strength,
their bargaining position. Of course, assuming such a bargaining
position will already bring additional elements into the game.
When, for instance, one of the players is in a position to sustain
higher losses, then their pay-offs are no longer  really equal,
because even though the numbers are the same, these
numbers matter less to some than to other players. \\
In any case, in a single-shot game without additional elements
that affect the players differently, there is no way to decide
between the different possible equilibria. This might change when
the game is played repeatedly. Everything else being equal, one
could, for instance, guess that in the long run, things will
average out, and the symmetric equilibrium \rf{4} might be
achieved.  Of course, playing the game repeatedly enlarges the
strategy space, in the sense that a player in a given round will
choose an action that may also affect her/his future pay-offs.
She/he may therefore try to induce the opponents to make
unfavorable, i.e.,  higher  contribution in subsequent rounds, for
instance by playing low her/himself, even though that might cause
failure in the present round. Of course, when the opponents are
doing the same, neither would gain any pay-off, and the question
then might be which of them can hold out longer. On the other
hand, from observing the opponents over a couple of rounds, a
player may also try to gain some insights into the opponents'
behavior and exploit observed regularities in that behavior. The
opponents, however, again will do the same. Thus, the players will
try to mutually adapt to each other. This will then trigger an
interesting dynamics when the number of rounds played  grows. The
players may utilize rather complex strategies then, for example
for making predictions of the opponents' future behavior on the
basis of observations of many rounds. In simulations, one may then
break the symmetry by allowing the players access to different
strategy types. In that way, one can investigate what type of
strategy is
better than others. \\
We can then also set up an evolutionary process on the basis of
some evaluation criterion. When we have populations of players, it
is natural to let each player not always play the same opponent,
but rather pair her/him with randomly selected opponents in each
round. In that way, the population gets sampled, and many effects
will get averaged out so that the essential structure of the game
will emerge. Again, each player is allowed to play a fixed number
of rounds, the same for everybody so that we can compare their
accumulated pay-offs. On the basis of that evaluation, the
evolutionary step then produces the next generation of players.
The more successful ones have more offspring than the others,
and the poor performers may not get any offspring at all.  \\
There are some sources of randomness involved. The members of the
populations are randomly initialized. For instance, in the
simplest case, each agent $a$ just can play one simple number
$k_a$ which might be randomly drawn from the integers between $0$
and $K$, independently and identically for each agent. Also, the
pairing of the agents from the different population in each round
is done randomly. Finally, the evolutionary step contains some
random selection and mutation scheme. In \cite{JostLiThird}, we
have systematically investigated the effects of the various
sources of
randomness. Therefore, here we shall not address this issue in detail.\\
Concerning the population, we have two options. We could either
work with one single  population in which $I$ players get randomly
grouped in each round, a game being successful when the sum of
their contributions is at least $K$, see \rf{1}, and receiving
pay-offs according to \rf{2}. Or we could work with  distinct
populations so that in each round, groups of $I$ players, one from
each population,
are randomly chosen to play the game. \\
In the first case, we could compose the -- single -- population of
different strategy types and see how they perform and how the
population evolves. In the second case, we could also equip the
populations with access to different strategy spaces and study how
those perform at the population level. In that case, we not only
have an -- indirect -- competition for accumulated pay-off within
each population, which decides about the evolutionary update, but
also a competition between the populations for achieving a
favorable equilibrium value. Thus, we can see the interaction of
competitions at two different scales, individuals struggling
inside their own population to be more successful than others, and
populations or
strategy types competing with each other at a cumulative scale. \\
In particular, we can investigate in a quantitative manner a
version of the tragedy of the commons. This is concerned with
situations where individual contributions are needed for a
collective benefit, but individuals can gain an advantage by
participating in the benefit without contributing. Others then
need to contribute if the benefit is to be sustained. Typically,
however, in the end nobody will contribute and therefore also
nobody will benefit. In our scenario, when some agents will stop
contributing, it is still advantageous for the other ones to
contribute more and fill the gap, as long as at least two agents
contribute. In turn, when only two contribute, neither of them can
gain an advantage by stopping her/his contributions. Thus, our
question here is under which circumstances some agents can
discover that by  contributing less, they will force the other
ones to contribute more. The problem for the agents is that when
at an equilibrium, one of them will suddenly decrease or terminate
her/his contributions, then all of them will loose, including the
defector. Therefore, the tendency to defect needs to start already
before the equilibrium is reached, and in turn it will then affect
that equilibrium. \\
Also, the fact that in a completely symmetric situation, some
agents might start to contribute less or stop contributing, is a
phenomenon of symmetry breaking due to random fluctuations which
can be investigated here in a simple setting. In particular, we
shall see that whereas the equilibrium reached for two or even
three populations tends to be symmetric, this is no longer the
case for more populations. Thus, in our scenario, the tragedy of
the commons results only when there are sufficiently many
participants.

\section{Defecting}
As explained in the introduction, in each instance of the game, we
form a group of $I$ players, one from each of the $I$ different
populations. To start with, we let each population have the same
number $N$ of agents. Therefore, in each round, we can let $N$
such disjoint groups play simultaneously. This ensures that every
player plays the same number of times. We then fix a generation
time $T$. That is, after playing $T$ rounds, the sum of the
pay-offs over these round is computed for each agent. In each
population, we then form a new generation by some evolutionary
scheme, that is, the expected number of  offspring in the next
generation is a positive function of an agent's accumulated
pay-off. The population size $N$ will be kept constant across
generations. Thus, successful agents could have more than one
offspring whereas less successful might not have any.  Thus, the
populations will evolve by rewarding
the more successful agents at the expense of the others within each population.\\
Here, we consider the simplest case where all populations have
access to the same strategy space. The question we want to address
then is whether the fair equilibrium \rf{4} is attained or not.
Our previous investigations \cite{JostLiFirst,JostLiSecond}
indicate that the equilibrium \rf{4} is stable for two
populations.\footnote{The game
  considered in these references uses two players (called ``buyers''
  and ``sellers'') that have to make
  offers $b, s$ between 0 and some maximum $K$. When $b \ge
  s$, the first player gets $K-b$, the second one $s$, else they get
  nothing.  Putting $k_1=b$, $k_2=K-s$, that game is converted into
  the one considered in the present paper.} Here, we investigate the
situation with 3 and more populations. \\
After sufficiently many generations, the game dynamics will
converge to some equilibrium where the players of each fixed
population always contribute the same offer $k_i$ such that
\bel{11} \sum_i k_i \ge K \qe (because of some random fluctuations
in the course of the dynamics, usually the sum is slightly larger
than $K$, that is, the populations use some safety margin in their
offers). Whenever the populations settle at such an equilibrium,
no single player can gain any advantage from deviating from it
(except for narrowing down the safety margin, but we shall ignore
that issue in our discussion). The equilibrium is a Nash
equilibrium (see e.g. \cite{Wei} for game theoretic concepts
utilized in our discussion). At the equilibrium, however, the
contributions $k_i$ coming from the different populations need not
all be equal. That is, inside each population, the agents are
homogeneous, but the populations themselves may be different from
each
other. \\
We find that, when we have 4 or more populations, the members of
some populations discover the possibility of defecting. This means
that the final equilibrium reached has at least one of the
$k_i=0$. Since the populations are symmetric to each other as
regards their sizes and strategy spaces, which of them discovers
the possibility of defecting is solely determined by random
fluctuations in the individual conditions or the dynamics. What we
are interested therefore is not
which population defects, but how many of them do. \\
For the resulting equilibrium, we find a minimal breaking of
symmetry. That is, after relabelling, there will be $I_1$
populations with $k_i=\frac{K}{I_1}$ for $1\le i \le I_1$ and
$I-I_1$ populations  with $k_j=0$ for $I_1 + 1\le j \le I$. That
is, $I-I_1$ populations defect completely, whereas the other ones
contribute equally.

\section{Simulations}

As already mentioned in the previous sections, here we utilize a
very simple strategy to study the defecting behavior when more
than two populations participate in the game. There are at least
two advantages by starting from a simple strategy. First, we can
identify the critical factors which may lead to defecting, if
available, more efficiently than we can when faced with a case of
a more complex strategy. Second, simple strategies usually may
allow for some analytical understanding, which is of course more
appealing. In this simple strategy, initially all the members of
each population choose random numbers uniformly distributed
between 0 and $K$. After certain rounds of interactions, say $T$,
their payoffs will be compared and the more successful players
will be more represented in the next generation, namely having
more offsprings. It is very probable that some poorly performed
players may not have any offsprings at certain and therefore will
be eliminated right away. To maintain the diversity of parents
pool, we allow random mutations to occurs, but only with a tiny
probability. In such a simple scheme one would expect that the
convergence to the
equilibrium is efficient when the population size is modest.\\
In our simulations, the random mutation rate is kept fixed to be 1
percent as usual. But other parameters such as the generation time
$T$, the population size $N$, the maximum offer $K$ and the number
of populations $I$ are not fixed. By varying those parameters we
simply intend to investigate how the defecting, once emerges,
depends on certain parameters. Or, how will the interplay between
the parameters may finally culture the selfish defecting behavior?
Which parameters are playing the leading role?\\
We have already known that the defecting does not appear in the
two-population version of our game. That is, each population has
to make certain contributions in order to reach a favorable goal.
It is not possible for defecting to occur under totally symmetric
situations. We have checked various possibilities of parameters'
combination as $I$, the number of populations, is increased up to
3, of course, under symmetric conditions. But the 3-population
game is very similar to the 2-population one. That is, there is
certainly no chance for defecting. In most cases, the equilibrium
ends up where each population shares the equal contribution.
\\
If $I$ is increased up to 4, some interesting phenomena show up.
Namely, we find that defecting, that is at least one population
contributes nothing, may emerge once certain critical condition,
regarding $T$, $N$ and $K$, is satisfied. The first requirement is
enforced upon the generation time $T$, which should be as small as
possible (better be 1 which corresponds to immediate comparison of
interaction outcomes). If $T$ is too large, the defecting tendency
will be suppressed as the accumulated pay-offs of the players who
intend to defect will be smaller than the counterparts of those
who contribute. Consequently the defecting tendency will be less
rewarded and eventually gets eliminated before the equilibrium is
reached. This also indicates that the tendency to defect needs to
start from the beginning and only takes effects if it is not
eliminated too early. The second requirement is related to both
$N$ and $K$. We find at a certain $K$, there is a critical
threshold of $N$ for the occurrence of defecting. Or equivalently
for a fixed $N$, there is a critical threshold of $K$. This
condition is quite natural as $N/K$ is proportional to the initial
number of defectors (those who offers 0) within a certain
population. This value has to be large enough so that the
defecting behavior can be spread within the system and eventually
dominates, which generates a uniform population whose members only
offers 0. This condition is reasonable as well if one makes a
connection between the epidemic spreading and the spreading of
defecting behavior
in our game.\\
In Fig. 1 we show one simulation of the 4-population game. We see
that at the start of the game the situation is nearly symmetric,
except for some fluctuations. But after some rounds of
interactions, one population starts to offer less than the rest of
populations do, probably due to some random fluctuations, though
rather small. Apparently that small chance grows very fast and
eventually the whole population contributes nothing whereas the
rest populations contribute almost equally. The 5-population game
is no more different than the 4-population one except that 2
populations may be defectors and the rest three are contributors
who could offer differently due to fluctuations accumulated.

\section{Conclusion}
We have investigated a game where several players need to
contribute enough so that the sum of the contributions reaches or
exceeds some given threshold. Each player thus hopes that she/he
can get away with a small contribution while the other players
contribute so much that the threshold is still reached. This is
the same situation for every player. Thus, the game is completely
symmetric. In particular, it possesses a symmetric Nash
equilibrium where all contribute equally. However, this is not the
only equilibrium. Any state where the sum of the contributions is
equal to the threshold is an equilibrium. This includes states
where some players do not contribute anything while the others
then have to make up the deficit. This is a version of the tragedy
of the commons. The question addressed in this contribution then
is under which circumstances defecting sets in an evolutionary
population game among equal agents. That is, when do the members
of some agent populations by chance, that is, as a result of
random fluctuations in the evolution of the game, discover that
they can contribute less and in turn force the others to
contribute more? We have found that under rather general
conditions, such a defecting behavior is found for 4 or more
players. That is, our evolutionary population game produces a
phase transition towards the tragedy of the commons.

\end{document}